\documentclass[11pt,preprint]{aastex}
\usepackage[]{natbib}

\slugcomment{Submitted to The Astrophysical Journal}

\shorttitle{Observations of Putative Pulsar in CTA~1}
\shortauthors{Halpern et al.}

\newcommand\asca{{\it ASCA\/}}
\newcommand\ro{{\it ROSAT\/}}
\newcommand\chandra{{\it Chandra}}
\newcommand\xmm{{\it XMM\/}-Newton}
\newcommand\rxj{RX~J0007.0+7303}
\newcommand\eg{3EG~J0010+7309}
\newcommand\pos{(J2000) $00^{\rm h}07^{\rm m}01.\!^{\rm s}56$, $+73^{\circ}03^{\prime}08.\!^{\prime\prime}1$}

\begin{document}

\title{X-ray, Radio, and Optical Observations of the Putative Pulsar
in the Supernova Remnant CTA 1}

\author{J. P. Halpern, E. V. Gotthelf, F. Camilo, D. J. Helfand}
\affil{Columbia Astrophysics Laboratory, Columbia University,
550 West 120th Street,\\ New York, NY 10027-6601}
\email{jules@astro.columbia.edu}
\author{S. M. Ransom}
\affil{Department of Physics, McGill University, Montreal,
QC H3A 2T8, Canada}

\begin{abstract}
A \chandra\ image of the central X-ray source \rxj\ in the supernova
remnant CTA~1 reveals a point source, a compact nebula, and a bent
jet, all of which are characteristic of energetic, rotation-powered
pulsars.
Using the MDM 2.4m telescope we obtain upper limits in the optical at
the position of the point source, \pos, determined to an accuracy of 
$0.\!^{\prime\prime}1$, of $B > 25.4$, $V > 24.8$, $R > 25.1$, and $I
> 23.7$; these correspond to an X-ray-to-optical flux ratio $\ga 100$.
Neither a VLA image at 1425~MHz, nor a deep pulsar search at 820~MHz
using the NRAO Green Bank Telescope, 
reveal a radio pulsar counterpart to an equivalent
luminosity limit at 1400~MHz of 0.02~mJy~kpc$^2$, which is
equal to the lowest luminosity known for any radio pulsar.
The \chandra\ point source accounts for $\approx 30\%$ of
the flux of \rxj, while its compact nebula plus jet comprise $\approx
70\%$.  The X-ray spectrum of the point source is fitted with a
power-law plus blackbody model with $\Gamma = 1.6 \pm 0.6$,
$kT_{\infty} = 0.13 \pm 0.05$~keV, and $R_{\infty} = 0.37$~km, values
typical of a young pulsar.  An upper limit of
$T^{\infty}_{\rm e} < 6.6 \times 10^5$~K
to the effective temperature of the entire neutron star surface is derived,
which is one of the most constraining data points on cooling models.
The 0.5--10~keV luminosity of \rxj\ is
$\approx 4 \times 10^{31}\ (d/1.4\ {\rm kpc})^2$ ergs~s$^{-1}$,
but the larger ($\sim 18^{\prime}$ diameter) synchrotron nebula
in which it is embedded is two orders of
magnitude more luminous. These properties allow us to estimate, albeit
crudely, that the spin-down luminosity of the underlying pulsar is in
the range $10^{36} - 10^{37}$ ergs~s$^{-1}$, and support
the identification of the high-energy $\gamma$-ray source \eg\
as a pulsar even though its spin parameters have not yet
been determined.

\end{abstract}


\section{Introduction}

The supernova remnant CTA~1 (G119.5+10.2) is a radio shell with a
diameter of $\approx 1\fdg8$ \citep*{sew95} and a center-filled
X-ray morphology.  It has a kinematic distance of
$d = 1.4 \pm 0.3$~kpc derived from an associated \ion{H}{1} shell
\citep{pin93}, and the remnant age from
a Sedov analysis is $t_S \approx 1.3 \times 10^4$~yr \citep{sla04}.  A
detailed X-ray spectral study of the SNR using \asca\ data showed that
the large-scale extended emission is likely of a synchrotron origin;
its total luminosity is $5.6 \times 10^{33}\ (d/1.4\ {\rm kpc})^2$
ergs~s$^{-1}$
in the 0.5--10 keV band \citep{sla97}.  The compact {\it ROSAT\/} PSPC
source \rxj\ is located in the brightest part of the synchrotron
nebula \citep{sew95,sla97,bra98}.  \rxj\ has the requisite properties
for a rotation-powered pulsar.  Its X-ray luminosity in the
0.1--2.4~keV band is $4.3 \times 10^{31}\ (d/1.4\ {\rm kpc})^2$ ergs~s$^{-1}$
\citep{sla97}, typical of young pulsars, but atypically small compared
to the total luminosity of the large synchrotron nebula in CTA~1.
Appealing to an empirical relation between X-ray luminosity and spin-down
power $\dot E$, \citet{sla97} argued that the current spin-down power of the
pulsar, required to account for the present nonthermal emission from
the entire synchrotron nebula,
is $\dot E = 1.7 \times 10^{36}$ ergs~s$^{-1}$.
Combining this with the Sedov age predicts a 
rotation period $P \la 0.17$~s and a dipolar surface magnetic field
strength $B_{\rm p} \la 6 \times 10^{12}$~G.

Within the boundary of CTA~1 lies \eg, one of the brighter unidentified
EGRET sources.  With a 95\% confidence localization diameter of only
$28^{\prime}$ and an intermediate Galactic latitude of $+10\fdg2$,
this EGRET/SNR coincidence is one of the most convincing among the
unproven identifications \citep{bra98}, especially
because of the likely pulsar in \rxj.  The $\gamma$-ray source itself
has all of the characteristics of the known pulsars.
It is not variable; other
$\gamma$-ray pulsars show little if any change in flux while most blazars, the
other major class of EGRET source, are often dramatically variable.
Its photon spectral index of $1.85 \pm 0.10$ \citep{har99} ($1.58
\pm 0.18$ between 70~MeV and 2~GeV, \citealt{bra98}), is similar to other
EGRET pulsars, and flatter than that of most blazars.  The estimated values of
$d, t_S, \dot E$, and $P$ are all typical of what one would expect for
a young $\gamma$-ray pulsar.  In particular, the distance to CTA~1 is
typical for the predicted EGRET pulsar population
\citep*[e.g.,][]{hal93,kaa96,rom95,yad95,yad97}, and the inferred
energetics are comparable to those of the Vela pulsar and other
sources with similar values of $\dot E$.

Recent observations with \xmm\ resolved \rxj\ into a point source and
a diffuse nebula, and found a two-component (blackbody plus power-law)
spectral fit that is consistent with a young pulsar \citep{sla04}.
Here, we present a higher resolution \chandra\ observation of these
features that further resolves out a prominent jet-like structure 
and possible evidence of a torus, features typical of pulsar wind
nebulae (PWNe) found around the most energetic of pulsars.
In \S 2 we present our \chandra\ imaging
and spectroscopic results, in \S 3 the optical observations, and in
\S 4 new radio imaging and deep radio search for pulsations
from \rxj.  In \S 5 we discuss the constraints on $\dot E$
for the putative pulsar in CTA~1 based on its X-ray and
(likely) $\gamma$-ray luminosities.
Throughout this paper, for clarity, we refer to the
small-scale nebula resolved by \chandra\ within \rxj\ as the
PWN, as distinct from the large
synchrotron nebula, which is also likely wind fed.

\section{\chandra\ Observation of \rxj }

The central region of CTA~1 was observered on 2003 April 13 with the
Advanced CCD Imaging Spectrometer \citep[ACIS;][]{bur97} onboard the
\chandra\ X-ray Observatory \citep*{wei96}.  The source \rxj\ was
positioned at the default location on the back-illuminated S3 chip of
the ACIS-S array.  The standard {\tt TIMED} readout with a frame time
of 3.2~s was used, and the data were
collected in {\tt VFAINT} mode.  A total of 50139~s of on-time was
accumulated, while the effective exposure live-time was 49484~s.  All
data reduction and analysis were performed using the CIAO (V3.0.1),
FTOOLS (V5.2), and XSPEC (V11.2) X-ray analysis software packages. No
time filtering was necessary as the background rate was stable over
the course of the observation.  Photon pile-up was not a consideration
as the total count rate in the point source of interest was $<
0.01$~s$^{-1}$.

We used the standard processed and filtered event data with the latest
aspect alignments, with the exception that the 0.5 pixel
($0.\!^{\prime\prime}25$) randomization that is ordinarily applied to
the photon positions was reversed, restoring slightly sharper images.
The $0.\!^{\prime\prime}5$ ACIS pixels slightly undersample the
on-axis point spread function of the \chandra\ mirrors in the restored
images, as the radius encircling $50\%$ of the energy is $\approx
0.\!^{\prime\prime}5$.  However, the spacecraft dithering and
time-dependent aspect solution preclude the need for additional
randomization, which degrades a study of faint point sources embedded
in diffuse emission.

\subsection{ACIS Image}

\begin{figure}
\epsscale{1.}
\vskip -1.0in
\plotone{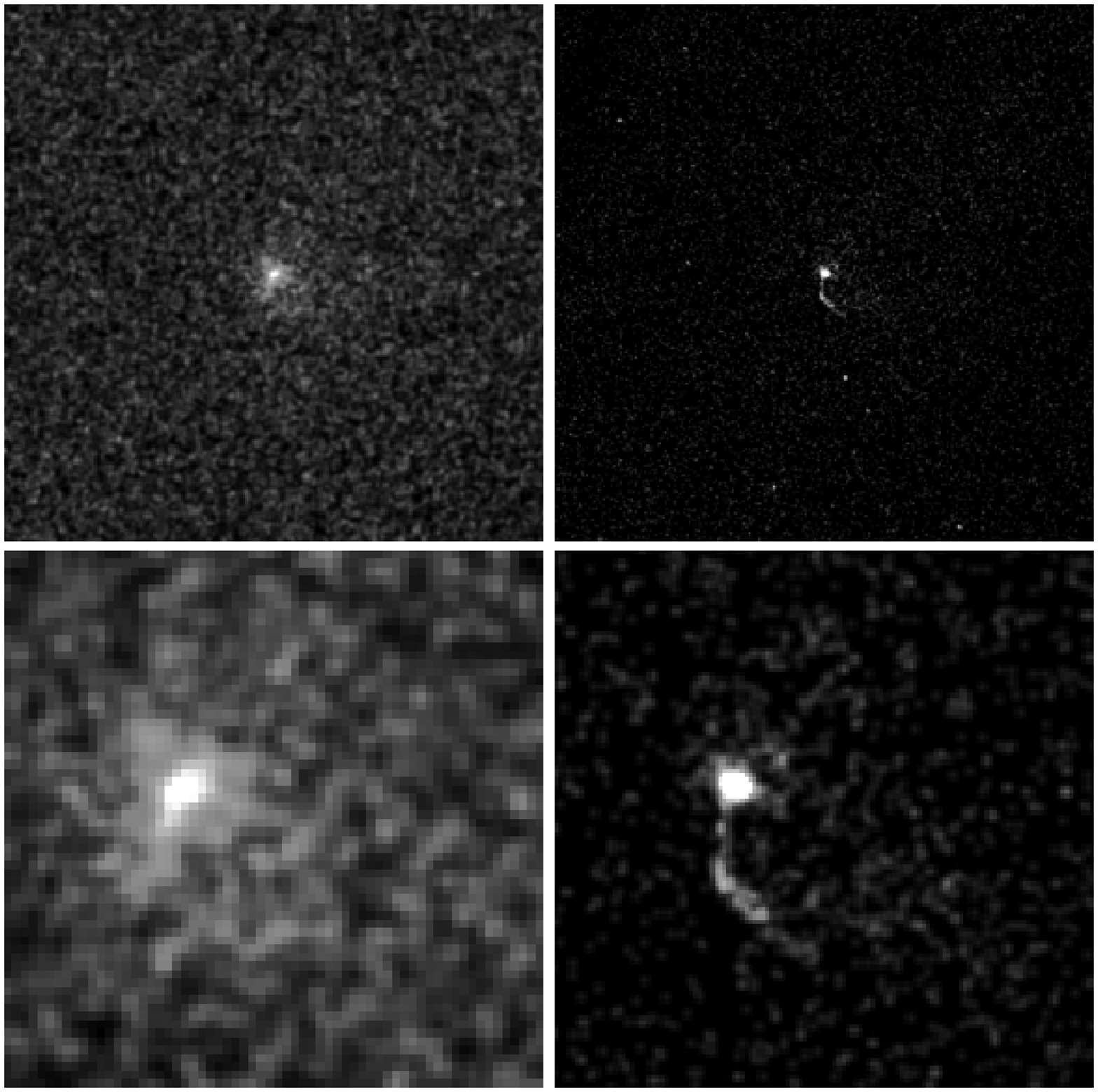}
\vskip -1.0in
\caption{\xmm\ and \chandra\ images of the central source \rxj\ in CTA~1
in the $0.5-8$~keV band, at two different display scales.  North
is up and east is to the left.
{\it Top} panels are $256^{\prime\prime}\times 256^{\prime\prime}$.
{\it Bottom} panels are $64^{\prime\prime}\times 64^{\prime\prime}$.
{\it Left}: Combined \xmm\ MOS1 and MOS2 images,
smoothed with a Gaussian of $\sigma = 1^{\prime\prime}$.
{\it Right}: \chandra\ ACIS-S3 image, smoothed with a Gaussian of
$\sigma = 0.\!^{\prime\prime}5$. The position of the point source is \pos.
[{\it A color version of this figure is available from the author.}]
\label{xray_image}}
\end{figure}

\begin{figure}
\includegraphics[width=4.3in,angle=270]{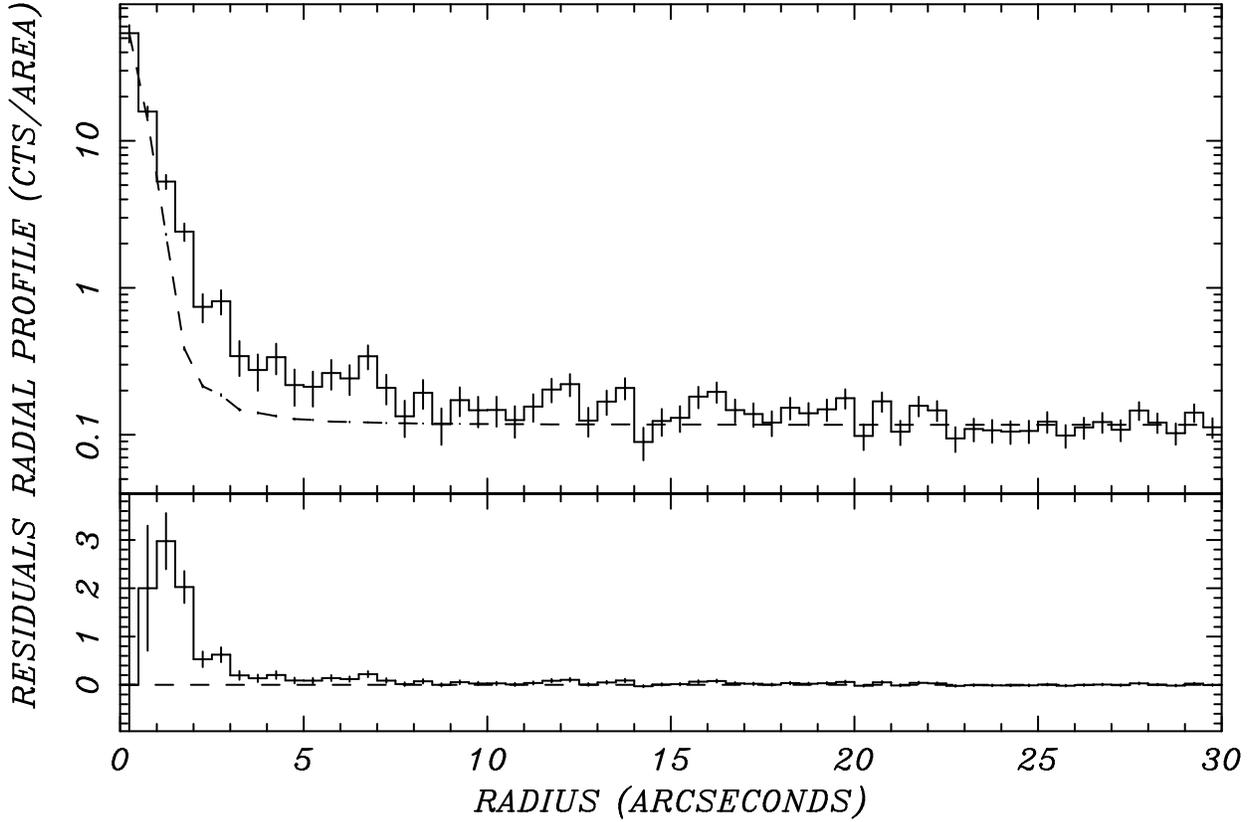}
\vskip 0.5in
\caption{\chandra\ radial profile of \rxj\ ({\it histogram})
compared to the PSF ({\it dashed line}),
which is calculated for the observed spectral distribution.
The PSF is scaled to match the detected counts in the central pixel
and the background measured in the annulus
$25^{\prime\prime} < r < 30^{\prime\prime}$.
A clear excess is seen corresponding to the faint nebulosity
at $r < 3^{\prime\prime}$, and additional enhancement
at $3^{\prime\prime} < r < 20^{\prime\prime}$
is likely due to the jet and fainter diffuse emission.
\label{radial_profile}}
\end{figure}

In Figure \ref{xray_image} we compare the \chandra\ ACIS image
of \rxj\ from 2003 April 13 with the combined MOS1 and MOS2 image
from the \xmm\ observation of 2002 February 21, originally reported by 
\citet{sla04}.
In the \chandra\ image, 187 photons (after background subtraction)
are found within $1^{\prime\prime}$ of the source maximum; most
are attributable to a point source.  
Figure~\ref{radial_profile} shows the radial profile of the \chandra\
source, with a simulated point-spread-function (PSF) scaled to the
detected counts in the central pixel.   There is also a
compact surrounding nebulosity with 136 photons within
a radius of $3^{\prime\prime}$ (the PWN),
as well as a jet that extends $16^{\prime\prime}$ to the south,
with a $\approx 50^{\circ}$ bend to the southwest at
$12^{\prime\prime}$ from the point source.  There is also a possible
faint extension of the jet for an additional $8^{\prime\prime}$ to the
southwest, although the latter may be part of a general expansion of
the jet into a large, low-surface brightness nebulosity extending to
the west and northwest.  The total number of photons in the jet is
45 after background subtraction.
From the Vela pulsar, similar diffuse emission apparently
supplied by its jet has been observed \citep{kar03,pav03}.

In view of {\it Chandra's} unique ability to reveal graphic evidence
of a rotation-powered neutron star, we refer in this paper to ``the
pulsar'' in \rxj\ rather than the customary circumlocution
``candidate pulsar.''  The ACIS frame time of 3.2~s does not permit
a search for X-ray pulsations in this observation. 
From the \xmm\ EPIC pn data taken with 6~ms time resolution,
\citet{sla04} derive an upper limit of 61\% sinusoidal
pulsed fraction for frequencies up to 83~Hz.
Nevertheless, we regard it as appropriate in cases such as
this one to classify an X-ray source as a pulsar even though its spin
parameters are not yet known.  The position of the pulsar is \pos\
(see \S 3 for a description of the astrometry).

\begin{figure}
\epsscale{1.}
\vskip -1.0in
\plotone{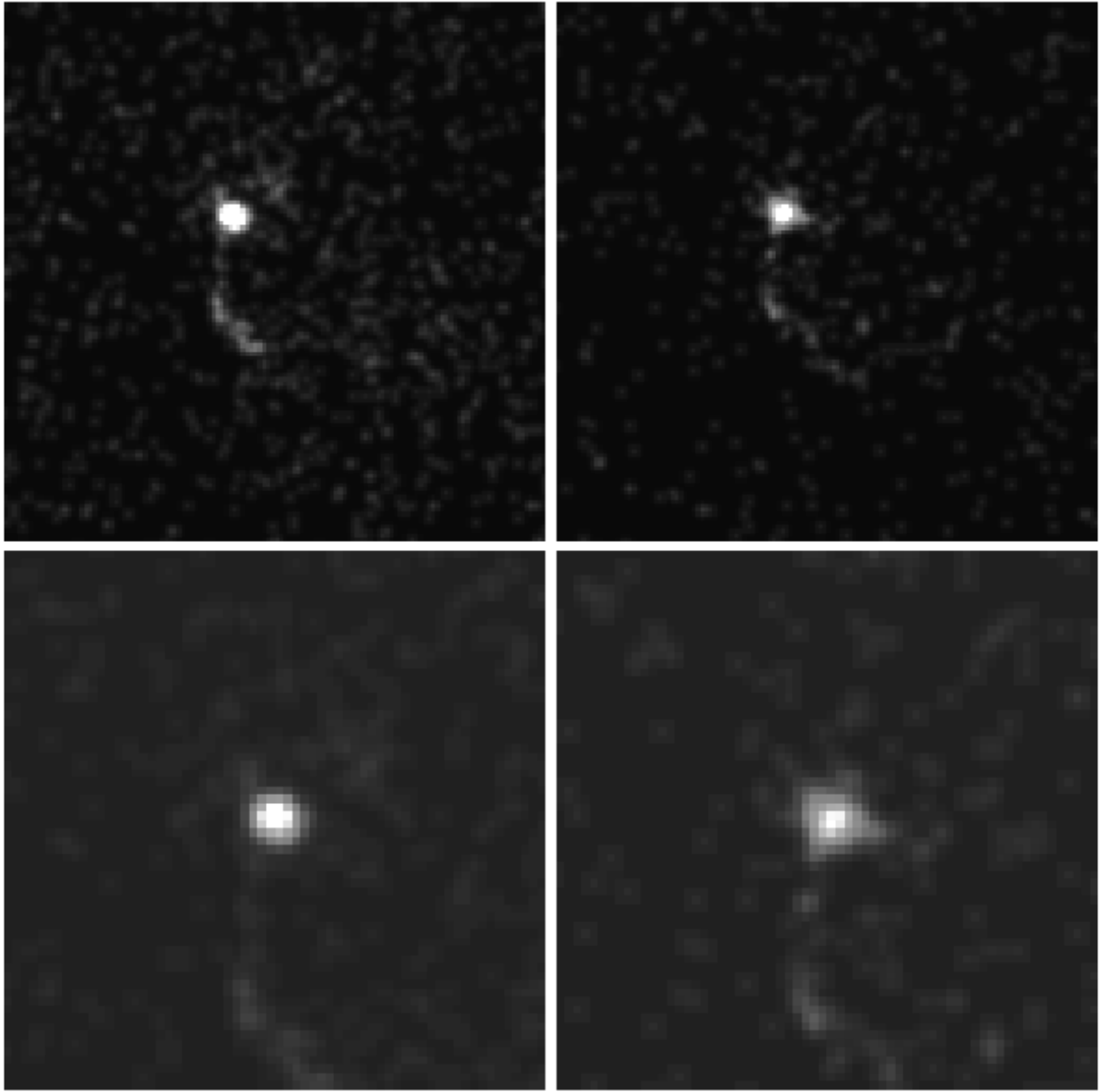}
\vskip -1.0in
\caption{\chandra\ images of the central source \rxj\ in CTA~1
in different energy bands and smoothed with a Gaussian of
$\sigma = 0.\!^{\prime\prime}5$. North is up and east is to the left.
{\it Left}: Soft (0.2--2.0~keV).
{\it Right}: Hard (2--8~keV).
{\it Top} panels are $64^{\prime\prime}\times 64^{\prime\prime}$.
{\it Bottom} panels are $32^{\prime\prime}\times 32^{\prime\prime}$.
[{\it A color version of this figure is available from the author.}]
\label{xray_color}}
\end{figure}

While the similarity between the
jets of \rxj\ and Vela is striking, it is curious that the prominent
torus structure of Vela is so much weaker in \rxj, if it is present at all.
The compact PWN surrounding the pulsar appears to be elongated
at an angle perpendicular to the inner jet, which by analogy
with Vela may plausibly be interpreted as an equatorial torus with the
jet emerging along the rotation axis.
Figure \ref{xray_color} shows the \chandra\ image in two different
energy bands, soft (0.2--2.0~keV) and hard (2--8 keV).  The PWN
is more prominent in the hard band, while the
pulsar itself is brighter in the soft band, presumably due to the
contribution of surface thermal emission from the neutron star (see \S
2.2.2), which has a softer spectrum than the synchrotron
nebulosity.  Apparent spectral differences exist along the length of the
jet.  The hardest regions of the jet are closest to the pulsar, while
the softest part of the jet is farther away, below the bend.

The features of the lower-resolution \xmm\ image are roughly
consistent with what one would expect from the \chandra\ image,
although the part of the jet past the bend is not obvious with \xmm.
It is possible that the structure of the jet has changed
markedly in the 14 months between the two observations.  Such behavior
would be consistent with the rapid changes seen in the jet of the Vela
pulsar \citep{kar03,pav03}, which moves with speeds in the range
$0.3-0.7\ c$, even in directions transverse to its length.
Velocities comparable to $c$ were also seen in \chandra\ observations
of G11.2--0.3 \citep{rob03}.  At the 1.4~kpc distance of CTA~1, the
$16^{\prime\prime}$ length of the jet corresponds to $3.4 \times
10^{17}$~cm, or 130 light days.  The length of the Vela pulsar's jet
is almost the same, $\approx 4.3 \times 10^{17}\ (d/300\ {\rm pc})$~cm,
which may suggest
that the two sources have similar spin-down luminosities.  The bending
and termination of the jet can plausibly be explained by its impact
against the larger synchrotron nebula.  In addition, if the pulsar is
traveling toward the southeast, the low surface brightness diffuse 
emission to the west and northwest of the pulsar may be due to jet
material dumping its energy and dispersing it into the larger
nebula, as suggested by \citet{pav03} in the case of Vela.

\subsection{X-ray Spectral Fitting}

Source and background spectra were extracted using the CIAO script
{\tt psextract} to generate the photon energy histograms and
appropriate response matrices for analysis of three spatial regions:
the pulsar ($r = 1^{\prime\prime}$), the PWN
($1^{\prime\prime} < r < 3^{\prime\prime}$), and the jet (irregular
polygon).  The PWN region served as background for the pulsar
spectrum, while a larger annular background region ($11^{\prime\prime}
< r < 15^{\prime\prime}$) was extracted for use with the PWN and the
jet spectra.  All spectra were grouped with a minimum of 20 counts per
bin.

\subsubsection{The PWN and its Intervening Column Density}

We first fitted the 136 background-subtracted counts obtained from the
PWN region to an absorbed power-law model. This produced a rather
unconstrained fit with $N_{\rm H} = 3.5 (0.0-8.9) \times
10^{21}$~cm$^{-2}$ and $\Gamma = 1.0 (0.3-1.4)$,
where the $1\sigma$ range is given in parentheses.
The unabsorbed flux in the $0.5-10$ keV band is $6.6 \times
10^{-14}$~ergs~cm$^{-2}$~s$^{-1}$ after adding a presumed
contribution, scaled by area, from within the pulsar region
($r = 1^{\prime\prime}$).  We note
that the best fitted value of $N_{\rm H} = 3.5 \times
10^{21}$~cm$^{-2}$ is somewhat larger than expected based on the total
Galactic optical extinction in the direction of CTA~1.  According to
the $100\,\mu$m IRAS/COBE maps \citep*{sch98}, $E(B-V) = 0.414$ mag,
corresponding to $A_V = 1.37$, which can be considered an upper limit
on the extinction to CTA~1.  The \citet{pre95} relation $N_H/A_V = 1.8
\times 10^{21}$ cm$^{-2}$ mag$^{-1}$ then implies $N_H = 2.5 \times
10^{21}$~cm$^{-2}$.  The latter value is close to that derived by
\citet{sla97} from joint fits to \ro\ PSPC and \asca\ spectra, which
gave $N_{\rm H} = 2.8 (2.3-3.4) \times 10^{21}$~cm$^{-2}$.
Given the distance and intermediate Galactic latitude of CTA~1
($+10\fdg2$), it is plausible that the source lies above the bulk of
the dust layer, since $1400\,{\rm sin}(10\fdg2) \approx 250$~pc.
Therefore, we adopt a fixed value of $N_{\rm H} = 2.8 \times
10^{21}$~cm$^{-2}$ in all further spectral fits, the results of which
are listed in Table~\ref{tbl-1}.

The 0.5--10 keV luminosity of the compact PWN region in \rxj\ is
$1.8 \times 10^{31}$~ergs~s$^{-1}$.  In comparison, the 
0.1--10 keV luminosity of the Vela nebula within a radius of
$53^{\prime\prime}$ ($2.4 \times 10^{17}$~cm) is
$(5-6) \times 10^{32}$~ergs~s$^{-1}$ \citep*{hel01,pav03},
a difference that is manifest in the relative prominence of
the toroidal features of the Vela nebula.

\subsubsection{The Pulsar}

We next fitted the pulsar spectrum, consisting of 187 background-subtracted
counts, to a two-component absorbed power-law plus blackbody model, to
allow for non-thermal magnetospheric emission and thermal emission from
the neutron star surface, respectively. The column density was held fixed
at $N_{\rm H} = 2.8 \times 10^{21}$~cm$^{-2}$. The best fitted
parameters are $\Gamma = 1.6 (1.1-2.2)$ and $kT_{\infty} = 0.13
(0.08-0.18)$~keV, with unabsorbed fluxes in the $0.5-10$ keV band of
$4.0 \times 10^{-14}$~ergs~cm$^{-2}$~s$^{-1}$ and $8.2 \times
10^{-15}$~ergs~cm$^{-2}$~s$^{-1}$, for the power-law and blackbody
components, respectively.  The small effective radius of the blackbody
component, $R_{\infty} = 0.37$~km, is similar to the value 
$R_{\infty} = 0.63$~km from the \xmm\ analysis \citep{sla04},
and might be evidence of a hot polar cap heated by backflow of energetic
particles.  On the other hand, \citet{sla04} also fitted a light-element
atmosphere model with a larger fixed radius of 10~km, enabling a lower
temperature to be derived.  Such systematic ambiguities are always
present in thermal modeling of neutron star spectra.

\begin{figure}
\epsscale{1.13}
\plottwo{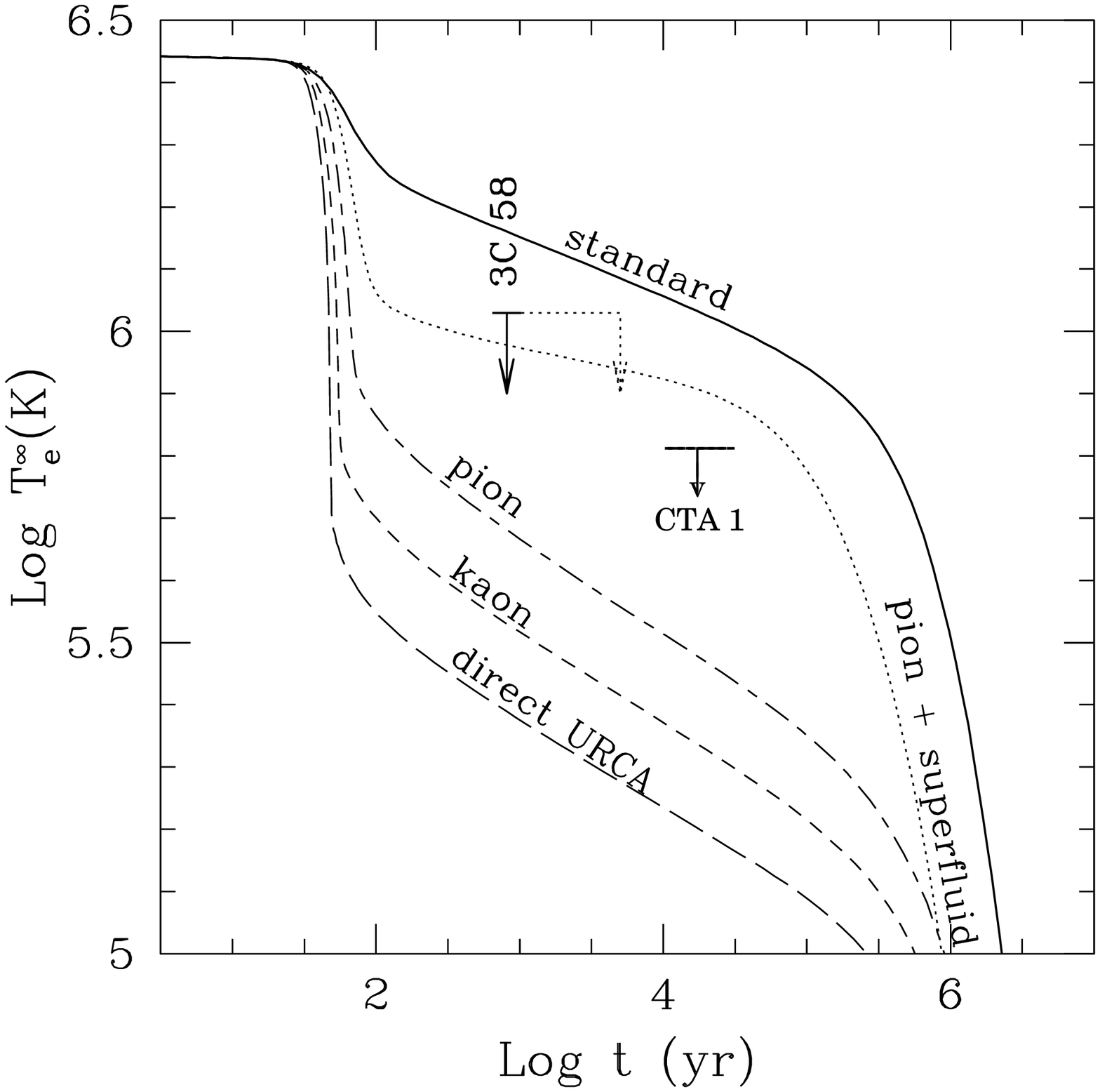}{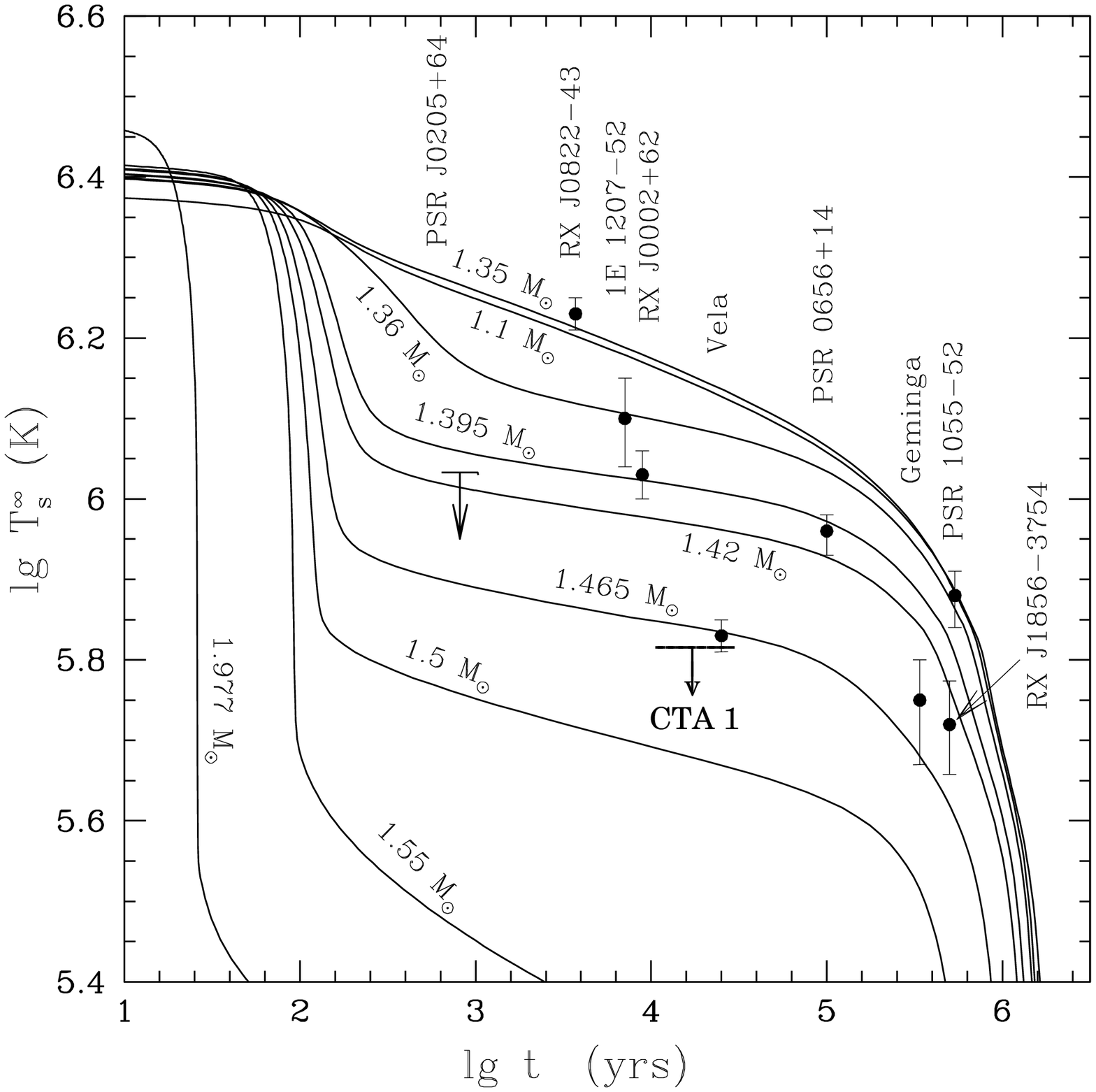}
\caption{{\it Left}: Upper limit on the surface temperature of the neutron
star in CTA1, in comparison with cooling models of \citet{pag98}.
The horizontal line denotes a possible age range of
$(1-3) \times 10^4$~yr.  This figure is reproduced from
\citet*{sla02}, and includes their upper limit on 3C~58.
{\it Right}: This figure is taken from \citet{yak02}, and includes
cooling models calculated assuming strong proton superfluidity
and weak neutron superfluidity.  Here, the direct Urca process is not
completely suppressed, and operates in the more massive neutron
stars.
\label{cooling_curves}}
\end{figure}

Independent of this spectral fit to the pulsar, we also derive an
upper limit to the effective blackbody temperature of the entire
neutron star surface.  Assuming $d=1.4$~kpc, $N_{\rm H} = 2.8 \times
10^{21}$~cm$^{-2}$, and a radius at infinity of 12~km, we compared
simulated blackbody spectra of increasing temperature with the data
until the predicted spectrum exceeded the observered counts in the
lowest energy bin by $3\sigma$, or nearly a factor of 2.  The resulting
upper limit is $T^{\infty}_{\rm e} < 6.6 \times 10^5$~K.
In Figure~\ref{cooling_curves}, we compare this value with a range of cooling
models that include both standard and ``exotic'' processes
(see also \citealt{pag92,tsu02}, and the reviews of neutron star cooling
by \citealt{tsu98} and \citealt{yak04}).  Recent calculations show
that the temperatures or upper limits on
all cooling neutron stars, such as Vela, Geminga,
3C~58 (PSR J0205+6449), and \rxj, can be fitted with models that
include proton and neutron superfluidity, and masses
$< 1.5\ M_{\odot}$ \citep*{kam02,yak02}, as shown in Figure~\ref{cooling_curves}.
Here, the direct Urca process is operating in the more massive neutron
stars.   Cooling can also be enhanced in a superfluid
neutron star by Cooper-pairing neutrino emission,
which was modeled most recently by \citet{gus04} and \citet{pag04}.

\subsubsection{The Jet}

For the jet we created a polygonal region which encompassed and followed
the contour of the jet from $2\farcs5$ to
$22^{\prime\prime}$ from the pulsar. A fit to the 45 background-subtracted 
photons with an absorbed power-law model, with $N_{\rm H}$ again fixed
to the \asca\ value, yielded $\Gamma = 1.26 (0.86-1.69)$ and an
unabsorbed flux of $3.1 \times 10^{-14}$~ergs~cm$^{-2}$~s$^{-1}$.
While there are probably changes in spectral index along the
jet that are related to the apparent differences in the soft and
hard images (Fig. \ref{xray_color}), there are too few photons
to fit spectra to multiple regions.  The 0.5--10 keV luminosity of
the jet, $7.3 \times 10^{30}$~ergs~s$^{-1}$, is almost identical to
that of the main (inner plus outer) jet of the Vela pulsar, which has
$L_X (1-8\ {\rm keV}) \approx 6 \times 10^{30}$~ergs~s$^{-1}$
\citep{pav03}.  Within the range of uncertainty and variability,
these two sources have jets of equal length and luminosity, which
is in curious contrast to the relative dominance of the toroids
in Vela.

\section{Optical Observations of \rxj }

\begin{figure}
\epsscale{1.}
\plotone{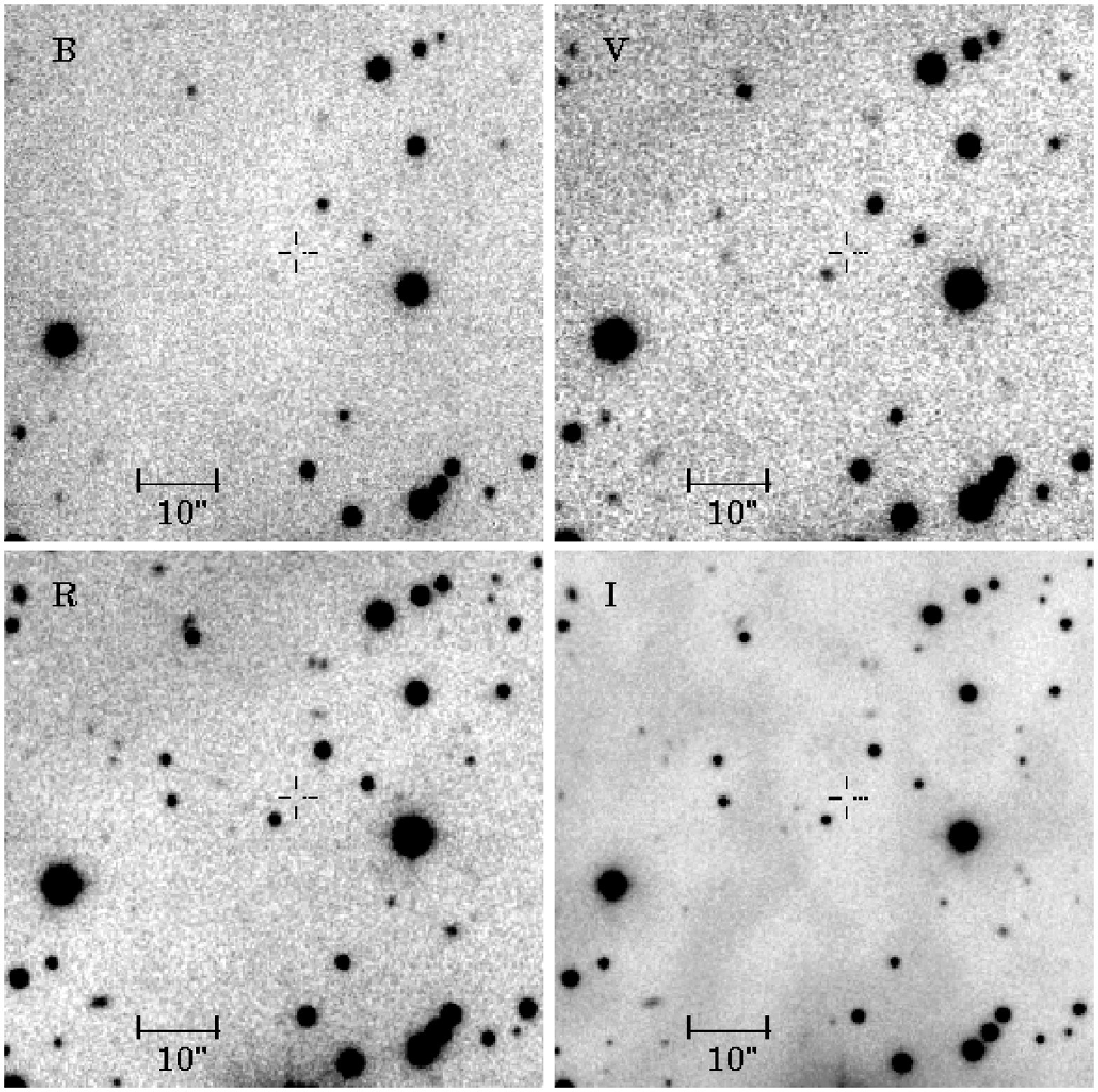}
\caption{CCD images at the location of the central X-ray source \rxj\
in CTA~1 obtained with the 2.4m Hiltner telescope.
North is up, and east is to the left.  Each panel
is $70^{\prime\prime}$ on a side.  Seeing ranges from
$0.\!^{\prime\prime}8$ in $I$ to $1.\!^{\prime\prime}2$ in $V$.
Interference fringes in the $I$-band image are artifacts.
Limiting ($3\sigma$) magnitudes at the position of
the \chandra\ point source, \pos\ as indicated by the {\it crosses},
are $B>25.4$, $V>24.8$, $R>25.1$, and $I>23.7$. 
[{\it A higher resolution version of this figure is available from the author.}]
\label{optical_image}}
\end{figure}

We obtained multiple CCD images of the field of \rxj\ using the 2.4m
Hiltner telescope of the MDM Observatory on 2002 August 26-30.  A
thinned, back-illuminated SITe $2048\times2048$ CCD with a spatial
scale of $0.\!^{\prime\prime}275$ per $24\,\mu$m pixel was used to
cover a $9.\!^{\prime}4 \times 9.\!^{\prime}4$ area. The sky
conditions were usually photometric enough to calibrate the $BVRI$
images with \citet{lan92} standard stars, although moonlight prevented
us from reaching the deepest possible limiting magnitudes.  Total
integration times were 100, 72, 120, and 75 minutes in $B, V, R$, and
$I$, respectively.  The seeing in the final, summed images, shown in
Figure~\ref{optical_image}, ranged from a best of
$0.\!^{\prime\prime}8$ in $I$ to a worst of $1.\!^{\prime\prime}2$ in
$V$. 

We were able to confirm and refine the astrometric accuracy of the
\chandra\ X-ray image relative to the optical reference frame of the
USNO-A2.0 catalog \citep{mon98} by identifying the counterparts of
several X-ray sources in the vicinity of \rxj.
X-ray positions and counts from the standard data processing,
and optical positions and magnitudes from our CCD images,
are listed in Table~\ref{tbl-2}.  The X-ray centroids and source counts
are calculated within the 80\% enclosed energy contour of the
local point-spread function.  All of the optical counterparts in
Table~\ref{tbl-2} appear point-like except the first entry,
CXOU J000647.2+730109, which is evidently a compact galaxy,
and perhaps the very faint CXOU J000739.2+730404, whose structure
is not clear.  The five neighboring sources had mean
X-ray--optical offsets of only $-0.\!^{\prime\prime}11$ in right
ascension and $+0.\!^{\prime\prime}06$ in declination, and rms scatter
of $0.\!^{\prime\prime}28$ in right ascension and
$0.\!^{\prime\prime}14$ in declination.  The rms scatter is consistent
with the statistical accuracies of the individual optical and X-ray
source positions, while the mean offset is well within the \chandra\
systematic 90\% confidence accuracy of $0.\!^{\prime\prime}6$.
Accordingly, we corrected the \chandra\ coordinates of the \rxj\ point
source by the small mean offsets, obtaining a final position of \pos,
which we expect is accurate to $\approx 0.\!^{\prime\prime}1$
relative to the optical reference frame.

Limiting ($3\sigma$) magnitudes at the position of the \chandra\
point source are $B>25.4$, $V>24.8$, $R>25.1$, and $I>23.7$.  The
extinction-corrected limits using the \citet{sch98} values of $A_B =
1.79, A_V = 1.37, A_R = 1.11$, and $A_I = 0.80$, are $B>23.6$,
$V>23.4$, $R>24.0$, and $I>22.9$.  This corresponds to
$f_X/f_V > 300$, or $>83$ after correction for extinction.  Similar
values are obtained in the other bands, and are consistent with an
isolated neutron star at the distance of CTA~1.  No extended optical
emission is seen associated with the X-ray jet.

\section{Radio Observations of \rxj }

\subsection{VLA Image}

\begin{figure}
\epsscale{1.}
\plotone{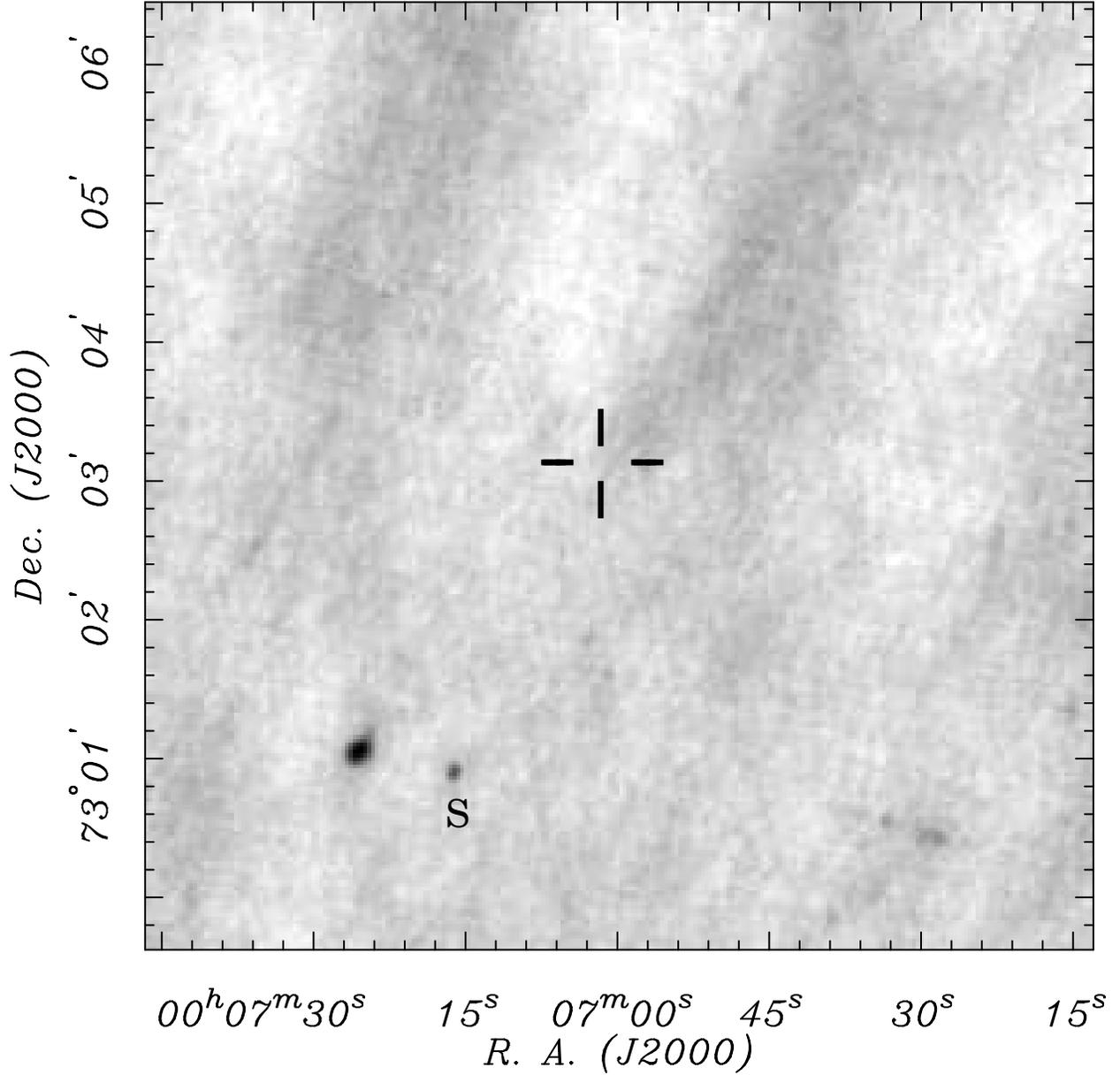}
\caption{VLA image of the central region of CTA~1 at 1425~MHz,
taken in AB configuration with a beam FWHM of
$5.\!^{\prime\prime}9 \times 4.\!^{\prime\prime}4$.
The cross marks the \chandra\ location of the neutron star in \rxj.
The $1\sigma$ noise level is $31\,\mu$Jy per beam
at this position.  For comparison, the point source marked ``S''
has a flux density of $180\,\mu$Jy.
The large-scale periodic noise 
in the image is an artifact.
\label{vla_image}}
\end{figure}

The field of \rxj\ was imaged at 1425 MHz using the AB configuration
of the VLA on 2002 May 27, which gave a beam FWHM of
$5.\!^{\prime\prime}9 \times 4.\!^{\prime\prime}4$.
An exposure time of 1~hour achieved a $1\sigma$
noise level of $31\,\mu$Jy and a non-detection at the
precise \chandra\ position of the pulsar.  This is equivalent to a
$3\sigma$ flux density upper limit of $0.09$~mJy.
The relevant part of the image in shown in Figure~\ref{vla_image}.

\subsection{Radio Pulsar Search at GBT}

CTA~1 was searched for pulsations by \citet{nic97} at radio
frequencies of 370~MHz and 1390~MHz.  The sensitivity of those
observations, translated to a standard frequency of 1400~MHz using a
spectral index $\alpha = -1.6$ (the median for all pulsars; \citealt{lor95}),
was $\sim 0.2$~mJy and $\sim 0.6$~mJy, respectively.  Also,
\citet*{lor98} reached a limit $S_{1400} \sim 0.2$~mJy
in their search at 600\,MHz.  Our VLA upper limit, 
$S_{1400} \la 0.1$~mJy, implies a luminosity $L_{1400} \equiv S_{1400}
d^2 \la 0.2$ mJy~kpc$^2$.  While this is a stringent limit (the
least luminous young radio pulsar known, in SNR 3C~58, has about twice
this luminosity; \citealt{cam02c}), there are weaker pulsars known.
We therefore used the extremely sensitive Green Bank Telescope (GBT),
where CTA~1 is a circumpolar source, to improve on this limit.

We observed the \chandra\ position of the pulsar
for a total of 19.60~hr on 2003 October~11.  We used
the Berkeley-Caltech Pulsar Machine (BCPM; \citealt{bac97}), an
analog/digital filter-bank with 96 channels for each of two polarizations,
to record a 48~MHz-wide band centered at a sky frequency of 820~MHz.
After summing in hardware the signals from the two polarizations,
removing the mean and scaling, the power levels from each frequency
channel were recorded with 4-bit precision to disk every $72\ \mu$s,
for a total of 980 million time samples.  [See \citet{cam02c} for
more details of a similar observation at the GBT using the BCPM.]

We analyzed the data using standard methods as implemented in the software
package PRESTO \citep{ran01}.  This involved identification and
removal of the worst radio-frequency interference (much of it generated at
the observatory), time-shifting the 96 frequency channels to compensate
for dispersive interstellar propagation prior to summing the respective
samples (doing this once for each assumed value of dispersion measure, the
integrated column density of free electrons), performing a Fast-Fourier
Transform of the resulting one-dimensional time series for each trial
DM after transformation to the solar-system barycenter, and sifting the
resulting spectrum for statistically significant peaks.  The last stage
included searching for narrow pulse shapes by summing up to eight harmonics
of the spectrum; and also searching for moderately accelerated signals,
as might be expected from a pulsar in a binary system, or a young one
with a large period derivative.

At a distance of 1.4~kpc, both the \cite{tay93} and \cite{cor02}
electron density models predict $\mbox{DM} \approx
25$~cm$^{-3}$~pc for the putative radio pulsar in CTA~1.  However,
the X-ray-measured $N_{\rm H} \approx 2.8\times10^{21}$~cm$^{-2}$
would predict $\mbox{DM} \approx 90$~cm$^{-3}$~pc, assuming an
average ratio $n_e/n_{\rm H} \approx 0.1$.  Also, the maximum Galactic
dispersion in the direction of CTA~1 predicted by the Cordes \& Lazio
model is 200~cm$^{-3}$~pc, so we searched up to this DM.  Although the
expected period for the pulsar in CTA~1 is $P \sim 0.1$~s, and almost
certainly not less than $\sim 20$~ms, we chose to analyze our data with
much higher time resolution in order to retain serendipitous sensitivity
to any pulsars located within the $15^{\prime}$-diameter GBT beam.  In order to
simplify the onerous reduction task, we first rebinned the data four-fold
to a time resolution of 0.288~ms, which we kept for the first 240 trial
DMs, each separated by 0.5~cm$^{-3}$~pc.  The effective time resolution
at the upper end of this range, limited by dispersive smearing within
one frequency channel, is about 1~ms.  We then further halved the time
resolution, doubled the DM step, and searched a further 80 time series.
In the end, we examined $\approx 100$ significant signals that appeared
at more than one DM, but none showed a true interstellar dispersion pattern
that would qualify it as a good pulsar candidate.

We now estimate the flux density limit of our observation.  The ideal rms
noise of the observation is $\sigma \equiv TG^{-1}(n B t)^{-1/2}$, where $T
\approx 35$~K includes the system temperature on cold sky as well as
a $\sim 7$~K contribution from Galactic synchrotron emission and $\sim
3$~K from CTA~1; $G = 2$~K~Jy$^{-1}$ is the telescope gain; $n = 2$
polarizations; $B$ is the bandwidth; and $t$ is the integration time.
With our parameters, $\sigma \approx 7\ \mu$Jy.  Including a $\sim 20\%$
factor accounting for losses due to hardware limitations and a threshold
signal-to-noise ratio of 8, we obtain an estimated sensitivity for a
sinusoidal pulse shape of $\sim 65\ \mu$Jy.  For a pulse with duty cycle
$w = 0.1\ P$, this limit is further improved by a factor of $\sim 3$.
Translating this to 1400~MHz with $\alpha \approx -1.6$, we obtain
$S_{1400} \la 30\ \mu$Jy for a sinusoidal pulse shape or $S_{1400} \la
10\ \mu$Jy for $w = 10\%$.  Strictly, these limits apply to relatively
long pulse periods; for $P \la 50$~ms the limits are somewhat worse,
the more so as $\mbox{DM}/P$ increases.

Taking $10\ \mu$Jy as the flux density limit at 1400~MHz for the pulsar
in CTA~1 implies $L_{1400} \la 0.02$ mJy~kpc$^2$.  This is equal to the
lowest luminosity known for a radio pulsar, and is a factor of $\approx 20$
below the least-luminous pulsar known with an age of less than 1~Myr.
If the pulsar in CTA~1 emits radio waves that intersect the Earth,
its luminosity is lower than that of any detected radio pulsar; perhaps
more likely, if it is a radio pulsar at all, its beam does not intersect
the Earth.

\section{Discussion: What is the Pulsar $\dot E$?}

The most important unknown property of \rxj\ is its spin-down
luminosity.  Predictions of $\dot E$ typically make use of
correlations between $L_X$ and $\dot E$ among known pulsars
\citep*[e.g.,][]{sew88,pos02}, but this method is fraught with
uncertainty, as well as special difficulties in the case of \rxj.  The
0.5--10~keV luminosity of \rxj\ that we measure with \chandra,
including the point and diffuse components, is $4.0 \times 10^{31}$
ergs~s$^{-1}$.  This is slightly smaller than the value of $5.4 \times
10^{31}$ ergs~s$^{-1}$ measured by \citet{sla04} using \xmm, the
difference probably attributable to an additional contribution of
diffuse emission contained in the larger \xmm\ extraction region. The
spin-down luminosity predicted by \citet{sla04} using the
\citet{sew88} relation is then $\dot E \approx 5 \times 10^{34}$ ergs~s$^{-1}$,
comparable to the values found for older $\gamma$-ray pulsars such as Geminga,
which has an $\dot E = 3.3 \times 10^{34}$ ergs~s$^{-1}$.
This interpretation is troubling in its neglect of
the large synchrotron nebula, for which there is no proposed power source 
other than the pulsar.  It is impossible to measure the entire synchrotron 
nebula of CTA~1 with \chandra\ because of its large size,
low surface brightness, and lack of a clear boundary.  But since the total 
synchrotron emission seen by \asca\ exceeds the compact source luminosity
by more than two orders of magnitude, consideration of the total energy
budget would favor a  minimum value of
$\dot E \approx 1.7 \times 10^{36}$ ergs~s$^{-1}$,
as originally proposed by \citet{sla97}.

We may also look to more detailed X-ray studies of pulsars with PWNe
to estimate the spin-down power of CTA~1. From an empirical perspective,
bright PWNe are found only around pulsars with $\dot E > 3 \times
10^{36}$ ergs~s$^{-1}$ \citep{got04}.\footnote{Extended emission recently 
found around the nearby ($d \sim 160$~pc) Geminga pulsar \citep{car03} 
has $L_X$ of only $\sim 6.5 \times 10^{28}$ ergs~s$^{-1}$,
a mere $2 \times 10^{-6}$ of its $\dot E$ or 2\% of its $L_X$.
Such weak emission would not have been detected at the distance of CTA~1.}
Above this threshold, PWN emission typically exceeds the pulsar
point-source luminosity by an order of magnitude.
Furthermore, the spectral index of the pulsar in the 2--10~keV band
correlates with $\dot E$ according to the \cite{got03} relation
$\Gamma_{\rm PSR} = 2.08 - 0.29\,\dot E_{38}^{-1/2}$, where $\dot E_{38}$
is the spin-down power in units of $10^{38}\ {\rm ergs\ s}^{-1}$ 
Applying this to CTA~1 for which
$\Gamma_{\rm PSR} = (1.1-2.2)$ suggests that $\dot E \ga 6 \times 10^{36}$
ergs~s$^{-1}$, allowing for scatter in the relation.

Further indication that $\dot E$ is in this higher range is suggested by the
$\gamma$-ray luminosity of \eg, which provides a lower-limit on the spin-down power.
At $d = 1.4$~kpc, the $\gamma$-ray flux integrated above 100~MeV corresponds
to $L_{\gamma} = 6 \times 10^{34}$ ergs~s$^{-1}$ if isotropic, or $5
\times 10^{33}$ ergs~s$^{-1}$ if beamed into 1~steradian as is often
(but arbitrarily) assumed.  Thus, a prediction of $\dot E \approx 5
\times 10^{34}$ ergs~s$^{-1}$ is incompatible with isotropic
$\gamma$-ray emission, and requires a substantial degree of beaming.
On the other hand, if $\dot E \approx 6 \times 10^{36}$
ergs~s$^{-1}$, then the ratio $L_{\gamma}/\dot E$ is only 0.01 in the
isotropic case, which is similar to Vela,
whose $\dot E = 6.9 \times
10^{36}$ ergs~s$^{-1}$.  Note that this argument is independent of
whether the X-ray spectrum of \rxj\ can be extrapolated over four
orders of magnitude in energy to join the $\gamma$-ray spectrum.  Such
a connection was suggested by \citet{sla04}, but it is impossible to
verify, as well as unnecessary for associating the sources.  The
$0.5-10$~keV power law of the Geminga pulsar clearly {\it does not\/}
connect with its $\gamma$-ray spectrum \citep{jac02}; rather, they are
separate spectral components arising from different emission mechanisms.

\section {Conclusions}

The resemblance of the X-ray morphology of \rxj\ to the PWN and jets
seen in \chandra\ images of the Crab \citep{wei02}, Vela
\citep{hel01,pav01,pav03}, PSR~B1509--58 \citep{gae02}, and other
newly discovered young pulsars \citep{hal01,lu02,cam02a,hes04},
enables us to point confidently to ``the pulsar'' in the SNR CTA~1
even though its spin parameters have not yet been determined.  Similar
conclusions were drawn from \chandra\ observations of supernova
remnants G0.9+0.1 \citep{gae01}, IC~443 \citep{olb01}, 3C 396
\citep{olb03}, and another unidentified EGRET source, GeV~J1809--2327
\citep{bra02}.  Using clues gathered from its
X-ray luminosity, morphology, spectrum, and likely
identification with \eg, we
predict that the spin-down power of the pulsar in \rxj\
is in the range $10^{36} - 10^{37}$ ergs~s$^{-1}$.

The upper limit of $T^{\infty}_{\rm e} < 6.6 \times 10^5$~K
that we derive on the effective 
temperature of the full neutron star surface in CTA~1
is more constraining of cooling models than nearly all
other cases. As described by \citet{yak02},
X-ray observations do not yet require exotic phases of matter
such as pion and kaon condensates, or free quarks.  However, 
it is necessary to allow a range of neutron star masses
if the measured temperatures and ages are to be accommodated
by a single equation of state and theory of superfluid
properties \citep{gus04}.  In this sense, the neutron star in CTA~1
is cooler {\it for its age} than most others, and similar to
Vela and Geminga in requiring a mass $> 1.42\ M_{\odot}$.
If so, it may become an issue to understand why
single neutron stars often have inferred masses greater than
the precise values that are measured in binary radio pulsars,
which cluster tightly around $1.35 \ M_{\odot}$
\citep{tho99,sta02,lyn04,wei03}.  Discussion of that potential
problem is beyond the scope of this paper.

Energetic pulsars are being discovered with extremely small radio
luminosities; whether this is intrinsic or due to beaming
remains unclear.  Recent examples
include PSR J2229+6114 coincident with 3EG~J2227+6122 \citep{hal01},
PSR J1930+1852 in G54.1+0.3 \citep{lu02,cam02a},  PSR J1124--5916
in G292.0+1.8 \citep{hug01,cam02b},
and PSR J0205+6449 in 3C~58 \citep{mur02,cam02c}.
All of these pulsars have $\dot E = (1-3)
\times 10^{37}$ ergs~s$^{-1}$ and luminosities at 1400~MHz of $\approx
1$ mJy~kpc$^2$.  Such radio luminosities are less than those of
$\approx 95\%$ of all pulsars, and less than every other pulsar
younger than $2 \times 10^4$~yr \citep{cam02b}.
In comparison, the luminosity upper limit for \rxj\ is
only $\approx 0.02$ mJy~kpc$^2$, which makes it fainter in the
radio than all pulsars except
the nearby Geminga and perhaps the putative pulsar powering 3EG~J1835+5918
\citep{mir00,mir01,hal02}. If due to unfavorable orientation with respect to
the radio beam, the likely existence of a radio-quiet $\gamma$-ray pulsar in
CTA~1 tends to support the outer-gap model, in which
the $\gamma$-rays arise from regions far out in the magnetosphere and
are emitted into a wide fan beam, illuminating a much larger fraction
of the sky than does the narrow radio beam.

\acknowledgments

We thank Don Backer for making the
BCPM available to the GBT user community.
This is work was supported by SAO grant GO3-4064X.

\clearpage

\clearpage

\begin{deluxetable}{lrrr}
\tablecaption{Spectral Fits to Chandra Observation of \rxj\label{tbl-1}}
\tablewidth{0pt}
\tablehead{
\colhead{Parameter\tablenotemark{a}} & \colhead{Pulsar} & \colhead{PWN} & \colhead{Jet}
}
\startdata
Net counts                          & 187     & 136       &       45 \\   
$N_{\rm H}$ (cm$^{-2}$) (fixed) & $2.8 \times 10^{21}$ & $2.8 \times 10^{21}$ & $2.8 \times 10^{21}$ \\
$\Gamma$                     & $1.62(1.10-2.20)$ & $0.97(0.61-1.35)$ & $1.26(0.86-1.69)$ \\
$C_{\rm pl}$ (photons cm$^{-2}$ s$^{-1}$ keV$^{-1}$)\tablenotemark{b} & $5.8 \times 10^{-6}$ &
$4.3 \times 10^{-6}$ & $2.9 \times 10^{-6}$\\
$F_{\rm pl}$ (ergs cm$^{-2}$ s$^{-1}$)\tablenotemark{c}    & $4.0 \times 10^{-14}$ & $7.5 \times 10^{-14}$&  $3.1 \times 10^{-14}$ \\
$L_{\rm pl}$ (ergs cm$^{-2}$ s$^{-1}$)\tablenotemark{c}    & $9.5 \times 10^{30}$  & $1.8 \times 10^{31}$& $7.3 \times 10^{30}$\\
$kT_{\infty}$ (keV)\tablenotemark{d}                       & $0.13(0.08-0.18)$ & . . . & . . . \\
$R_{\infty}$ (km)\tablenotemark{d}                         & $0.37$            & . . . & . . . \\
$F_{\rm bb}$ (ergs cm$^{-2}$ s$^{-1}$)\tablenotemark{c} & $8.2 \times 10^{-15}$ & . . . & . . . \\ 
$L_{\infty}$(bol) (ergs cm$^{-2}$ s$^{-1}$)\tablenotemark{d} & $5.0 \times 10^{30}$ &  . . . & . . . \\
$\chi^2_{\nu}$[dof] 					    & 0.58[6] &  0.68[4]  &  0.75[5]  \\
\enddata
\tablenotetext{a}{Uncertainties are 68\% confidence for one interesting parameter.}
\tablenotetext{b}{Power-law normalization at 1 keV.}
\tablenotetext{c}{Unabsorbed fluxes and luminosities in the 0.5--10 keV band,
assuming $d = 1.4$~kpc.}
\tablenotetext{d}{Spherical blackbody parameters for $d=1.4$~kpc.}
\end{deluxetable}

\clearpage
\begin{deluxetable}{lcrcccrr}
\tabletypesize{\scriptsize}
\rotate
\tablecaption{\chandra\ Sources and Optical Counterparts \label{tbl-2}}
\tablewidth{0pt}
\tablehead{
\colhead{Source} &
\colhead{X-ray Position} &
\colhead{Photons} &
\colhead{Optical Position} &
\colhead{$\Delta$ R.A.} &
\colhead{$\Delta$ Decl.} & 
\colhead{$B$} &
\colhead{$R$} \\
\colhead{} &
\colhead{R.A. \hskip 0.4in  Decl.} &
\colhead{(0.1--10 keV)} &
\colhead{R.A. \hskip 0.4in  Decl.} &
\colhead{($^{\prime\prime}$)} &
\colhead{($^{\prime\prime}$)} &
\colhead{(mag)} &
\colhead{(mag)}
}
\startdata
CXOU J000647.2+730109 & 00 06 47.244 +73 01 09.12 &  15\phantom{XXX} & 00 06 47.369 +73 01 09.33  & $-0.55$ & $-0.21$ & 24.8\phantom{X} & $22.7$\phantom{X} \\
CXOU J000659.4+730219 & 00 06 59.414 +73 02 19.16 &  23\phantom{XXX} & 00 06 59.393 +73 02 19.00  & $+0.09$ & $+0.16$ & 20.3\phantom{X} & $18.4$\phantom{X} \\
\rxj\                 & 00 07 01.537 +73 03 08.17 & 274\phantom{XXX} &            . . .           & 
. . .  & . . .   & $>25.4$\phantom{X}  & $>25.1$\phantom{X} \\
CXOU J000739.2+730404 & 00 07 39.276 +73 04 04.74 &  36\phantom{XXX} & 00 07 39.289 +73 04 04.58  & $-0.06$ & $+0.16$ & 25.2\phantom{X} & $23.6$\phantom{X}  \\
CXOU J000740.3+730649 & 00 07 40.343 +73 06 49.43 & 176\phantom{XXX} & 00 07 40.408 +73 06 49.31  & $-0.28$ & $+0.12$ & 23.3\phantom{X} & $21.6$\phantom{X}  \\
CXOU J000741.9+730227 & 00 07 41.936 +73 02 27.17 & 158\phantom{XXX} & 00 07 41.877 +73 02 27.11  & $+0.26$ & $+0.06$ & 23.6\phantom{X} & $22.1$\phantom{X} \\
\enddata
\tablecomments{Units of right ascension are hours, minutes, and seconds,
and units of declination are degrees, arcminutes, and arcseconds.}
\end{deluxetable}


\begin{thebibliography}{}

\bibitem[Backer et al.(1997)]{bac97}
Backer, D. C., Dexter, M. R., Zepka, A., Ng, D., Werthimer, D. J., Ray, P. S.,
\& Foster, R. S. 1997, \pasp, 109, 61

\bibitem[Braje et al.(2002)]{bra02}
Braje, T. M., Romani, R. W., Roberts, M. S. E., \& Kawai, N. 2002, \apjl, 565, L91

\bibitem[Brazier et al.(1998)]{bra98}
Brazier, K. T. S., Reimer, O., Kanbach, G., \& Carrami\~nana, A. 1998, \mnras, 295, 819

\bibitem[Burke et al.(1997)]{bur97} Burke, B. E., Gregory, J.,
Bautz, M. W., Prigozhin, G. Y., Kissel, S. E., Kosicki, B. N.,
Loomis, A. H., \& Young, D. J. 1997, IEEE Trans. Electron Devices,
44, 1633

\bibitem[Camilo et al.(2002a)]{cam02a} Camilo, F., Lorimer, D. R.,
Bhat, N. D. R., Gotthelf, E. V., Halpern, J. P., Wang, Q. D., Lu, F. J.,
\& Mirabal, N. 2002a, \apjl, 574, L71

\bibitem[Camilo et al.(2002b)]{cam02b} Camilo, F., Manchester, R. N.,
Gaensler, B. M., Lorimer, D. R., \& Sarkissian, J. 2002b, \apjl, 567, L71

\bibitem[Camilo et al.(2002c)]{cam02c} Camilo, F., et al. 2002c, \apjl, 571, L41

\bibitem[Caraveo et al.(2003)]{car03} Caraveo, P. A., Bignami, G. F., DeLuca, A., Mereghetti, S., Pellizzoni, A., Mignani, R., Tur, A., \& Becker, W. 2003,
Science, 301, 1345

\bibitem[Cordes \& Lazio(2002)]{cor02}
Cordes, J. M., \& Lazio, T. J. W. 2002, preprint (astro-ph/0207156)

\bibitem[Gaensler et al.(2002)]{gae02}
Gaensler, B. M., Arons, J., Kaspi, V. M., Pivovaroff, M. J., Kawai, N.,
\& Tamura, K. 2002, \apj, 569, 878

\bibitem[Gaensler, Pivovaroff, \& Garmire(2001)]{gae01}
Gaensler, B. M., Pivovaroff, M. J., \& Garmire, G. P. 2001, \apj, 556, L107

\bibitem[Gotthelf(2003)]{got03} Gotthelf, E. V. 2003, \apj, 591, 361

\bibitem[Gotthelf(2004)]{got04} Gotthelf, E. V. 2004, in IAU Symp. 218,
Young Neutron Stars and their Environment,
ed. F. Camilo \& B. M. Gaensler (San Francisco: ASP), 000

\bibitem[Gusakov et al.(2004)]{gus04} Gusakov, M. E., Kaminker, A. D.,
Yakovlev, D. G., \& Gnedin, O. Y. 2004, \aap, submitted (astro-ph/0404002)

\bibitem[Halpern et al.(2001)]{hal01}
Halpern, J. P., Camilo, F., Gotthelf, E. V., Helfand, D. J., Kramer, M.,
Lyne, A. G., Leighly, K. M., Eracleous, M. 2001, \apj, 552, L125

\bibitem[Halpern et al.(2002)]{hal02} Halpern, J. P., Gotthelf, E. V.,
Mirabal, N., \& Camilo. F. 2002, \apjl, 573, L41

\bibitem[Halpern \& Ruderman(1993)]{hal93}
Halpern, J. P., \& Ruderman, M. 1993, \apj, 415, 286

\bibitem[Hartman et al.(1999)]{har99}
Hartman, R. C., et al. 1999, \apjs, 123, 79

\bibitem[Helfand et al.(2001)Helfand, Gotthelf, \& Halpern]{hel01}
Helfand, D. J., Gotthelf, E. V., \& Halpern J. P. 2001, \apj, 556, 380

\bibitem[Hughes et al.(2001)]{hug01}
Hughes, J. P., Slane, P. O., Burrows, D. N., Garmire, G., Nousek, J. A.,
Olbert, C. M., \& Keohane, J. W. 2001, \apj, 559, L153

\bibitem[Hessels et al.(2004)]{hes04} Hessels, J. W. T., Roberts, M. S. E.,
Ransom, S. M., Kaspi, V. M., Romani, R. W., Ng, C.-Y., Freire, P. C. C.,
\& Gaensler, B. M. 2004, \apj, in press (astro-ph/0403632)

\bibitem[Jackson et al.(2002)]{jac02} Jackson, M. S., Halpern, J. P.,
Gotthelf, E. V., \& Mattox, J. R. 2002, \apj, 578, 935

\bibitem[Kaaret \& Cottam(1996)]{kaa96}
Kaaret, P., \& Cottam, J. 1996, \apj, 462, L1

\bibitem[Kaminker et al.(2002)Kaminker, Yakovlev, \& Gnedin]{kam02}
Kaminker, A. D., Yakovlev, D. G., \& Gnedin, O. Y. 2002, \aap, 383, 1076

\bibitem[Kargaltsev et al.(2003)]{kar03}
Kargaltsev, O., Pavlov, G. G., Teter, M. A., \& Sanwal, D. 2003, NewAR, 47, 487

\bibitem[Landolt(1992)]{lan92} Landolt, A. U. 1992, \aj, 104, 340

\bibitem[Lorimer et al.(1998)Lorimer, Lyne, \& Camilo]{lor98}
Lorimer, D. R., Lyne, A. G., \& Camilo, F. 1998, \aap, 331, 1002

\bibitem[Lorimer et al.(1995)]{lor95}
Lorimer, D. R., Yates, J. A., Lyne, A. G., \& Gould, D. M. 1995, \mnras, 273, 411

\bibitem[Lu et al.(2002)]{lu02} Lu, F. J., Wang, Q. D., Aschenbach, B.,
Durochoux, P., \& Song, L. M. 2002, \apj, 568, L49

\bibitem[Lyne et al.(2004)]{lyn04} Lyne, A. G., et al. 2004, Science, 303, 1153

\bibitem[Mirabal \& Halpern(2001)]{mir01} Mirabal, N., \& Halpern, J. P.
2001, \apjl, 547, L137

\bibitem[Mirabal et al.(2000)]{mir00} Mirabal, N., Halpern, J. P.,
Eracleous, M., \& Becker, R. H. 2000, \apj, 541, 180

\bibitem[Monet et al.(1998)]{mon98}
Monet, D., et al. 1998, USNO-SA2.0 (Washington, DC: US Naval Obs.)

\bibitem[Murray et al.(2002)]{mur02} Murray, S. S., Slane, P. O., Seward, F. D.,
Ransom, S. M., \& Gaensler, B. M. 2002, \apj, 568, 226

\bibitem[Nice \& Sayer(1997)]{nic97}
Nice, D. J., \& Sayer, R. W. 1997, \apj, 476, 261

\bibitem[Olbert et al.(2001)]{olb01}
Olbert, C. M., Clearfield, C. R., Williams, N. E., Keohane, J. W., \& Frail, 
D. A. 2001, \apjl, 554, L205

\bibitem[Olbert et al.(2003)]{olb03}
Olbert, C. M., Keohane, J. W., Arnaud, K. A., Dyer, K. K., Reynolds, S. P.,
\& Safi-Harb, S. 2003, \apjl, 592, L450

\bibitem[Page (1998)]{pag98} Page, D. 1998 in The Many Faces of Neutron Stars,
ed. R. Buccheri, J. van Paradijs, \& M. A. Alpar (Dordrecht: Kluwer), 539

\bibitem[Page \& Applegate(1992)]{pag92} Page, D., \& Applegate, J. H. 1992,
\apjl, 392, L17

\bibitem[Page et al.(2004)]{pag04} Page, D., Lattimer, J. M., Prakash, M.,
\& Steiner, A. W. 2004, \apj, submitted (astro-ph/0403657)

\bibitem[Pavlov et al.(2001)]{pav01}
Pavlov, G. G., Kargaltsev, O. Y., Sanwal, D., \& Garmire, G. P. 2001, \apjl, 554, L189

\bibitem[Pavlov et al.(2003)]{pav03}
Pavlov, G. G., Teter, M. A., Kargaltsev, O., \& Sanwal, D. 2003, \apj, 591, 1157

\bibitem[Pineault et al.(1993)]{pin93}
Pineault, S., Landecker, T. L., Madore, B., \& Gaumont-Guay, S. 1993, \aj, 105, 1060

\bibitem[Possenti et al.(2002)]{pos02}
Possenti, A., Cerutti, R., Colpi, M., \& Mereghetti, S. 2002, \aap, 387, 993

\bibitem[Predehl \& Schmitt(1995)]{pre95}
Predehl, P., \& Schmitt, J. H. M. M. 1995, \aap, 293, 889

\bibitem[Ransom (2001)]{ran01} Ransom, S. M. 2001, Ph.D. thesis, Harvard Univ.

\bibitem[Roberts et al.(2003)]{rob03} Roberts, M. S. E., Tam, C. R.,
Kaspi, V. M., Lyutikov, M., Vasisht, G., Pivovaroff, M., Gotthelf, E. V.,
\& Kawai, N. 2003, \apj, 588, 992

\bibitem[Romani \& Yadigaroglu(1995)]{rom95}
Romani, R. W., \& Yadigaroglu, I.-A. 1995, \apj, 438, 314

\bibitem[Schlegel et al.(1998)Schlegel, Finkbeiner, \& Davis]{sch98}
Schlegel, D. J., Finkbeiner, D. P., \& Davis, M. 1998, \apj, 500, 525

\bibitem[Seward \& Wang(1988)]{sew88} Seward, F. D., \& Wang, Z.-R. 1988, \apj, 332, 199

\bibitem[Seward et al.(1995)Seward, Schmidt, \& Slane]{sew95}
Seward, F. D., Schmidt, B., \& Slane, P. 1995, \apj, 453, 284

\bibitem[Slane, et al.(2002)Slane, Helfand, \& Murray]{sla02}
Slane, P., Helfand, D. J., \& Murray, S. S. 2002, \apj, 571, L45

\bibitem[Slane et al.(1997)]{sla97}
Slane, P., Seward, F., Bandiera, R., Torii, K., \& Tsunemi, H. 1997, \apj, 485, 221

\bibitem[Slane et al.(2004)]{sla04} Slane, P., Zimmerman, E. R., Hughes, J. P., Seward, F. D., Gaensler, B. M., \& Clarke, M. J. 2004, \apj, 601, 1045

\bibitem[Stairs et al.(2002)]{sta02} Stairs, I. H., Thorsett, S. E.,
Taylor, J. H., \& Wolszczan, A. 2002, \apj, 581, 501

\bibitem[Taylor \& Cordes(1993)]{tay93}
Taylor, J. H., \& Cordes, J. H. 1993, \apj, 411, 674

\bibitem[Thorsett \& Chakrabarty(1999)]{tho99}
Thorsett, S. E., \& Chakrabarty, D. 1999, \apj, 512, 288

\bibitem[Tsuruta(1998)]{tsu98} Tsuruta, S. 1998, Phys. Rep., 292, 1

\bibitem[Tsuruta et al.(2002)]{tsu02} Tsuruta, S., Teter, M. A.,
Takatsuka, T., Tatsumi, T., \& Tamagaki, R. 2002, \apjl, 571, L143

\bibitem[Weisberg \& Taylor(2003)]{wei03}
Weisberg, J. M., \& Taylor, J. H. 2003, in ASP Conf. Proc. 302,
Radio Pulsars, ed. M. Bailes, D. J. Nice, \& S. E. Thorsett
(San Francisco: ASP), 93

\bibitem[Weisskopf et al.(1996)Weisskopf, O'Dell, \& van Speybroeck]{wei96} 
Weisskopf, M. C., O'Dell, S. L., \& van Speybroeck, L. P. 1996, Proc. SPIE, 
2805, 2

\bibitem[Weisskopf et al.(2000)]{wei02} Weisskopf, M. C., et al. 2000,
\apjl, 536, L81

\bibitem[Yadigaroglu \& Romani(1995)]{yad95}
Yadigaroglu, I.-A., \& Romani, R. W. 1995, \apj, 449, 211

\bibitem[Yadigaroglu \& Romani(1997)]{yad97} 
---------. 1997, \apj, 476, 347

\bibitem[Yakovlev \& Pethick(2004)]{yak04} Yakovlev, D. G., \& Pethick, C. J.
2004, \araa, in press (astro-ph/0402143)

\bibitem[Yakovlev et al.(2002)]{yak02} Yakovlev, D. G., Kaminker, A. D.,
Haensel, P., \& Gnedin, O. Y. 2002, \aap, 389, L24

\end{thebibliography}
\end{document}